\documentclass[sigconf,9pt]{acmart}
\renewcommand\footnotetextcopyrightpermission[1]{}
\usepackage{booktabs}
\usepackage{color}
\usepackage{graphicx}
\usepackage{url}
\usepackage{multirow}

\usepackage{xspace}

\usepackage{subfigure}
\usepackage{amsmath,chemformula}

\usepackage{pbox}
\usepackage{threeparttable,booktabs,multicol,multirow}
\setcopyright{none}

\begin{document}
\fancyhead{}

\title{Design Space Exploration for PCM-based Photonic Memory
}



\author{Amin Shafiee}
\affiliation{%
  \institution{Colorado State University}
  \city{Fort Collins}
  \country{USA}}

\author{Benoit Charbonnier}
\affiliation{%
  \institution{Univ. Grenoble Alpes, CEA, LETI}
  \city{F38000 Grenoble}
  \country{France}}


\author{Sudeep Pasricha}
\affiliation{%
  \institution{Colorado State University}
  \city{Fort Collins}
  \country{USA}}

\author{Mahdi Nikdast}
\affiliation{%
  \institution{Colorado State University}
  \city{Fort Collins}
  \country{USA}}

\begin{abstract}

The integration of silicon photonics (SiPh) and phase change materials (PCMs) has created a unique opportunity to realize adaptable and reconfigurable photonic systems. In particular, the nonvolatile programmability in PCMs has made them a promising candidate for implementing optical memory systems. In this paper, we describe the design of an optical memory cell based on PCMs while exploring the design space of the cell in terms of PCM material choice (e.g., GST, GSST, Sb$_2$Se$_3$), cell bit capacity, latency, and power consumption. Leveraging this design-space exploration for the design of efficient optical memory cells, we present the design and implementation of an optical memory array and explore its scalability and power consumption when using different optical memory cells. We also identify performance bottlenecks that need to be alleviated to further scale optical memory arrays with competitive latency and energy consumption, compared to their electronic counterparts.
 
\end{abstract}



 \maketitle
\pagestyle{empty} 

\section{Introduction}

\begin{figure*}[t]
    \centering
    \includegraphics[width=0.9\textwidth]{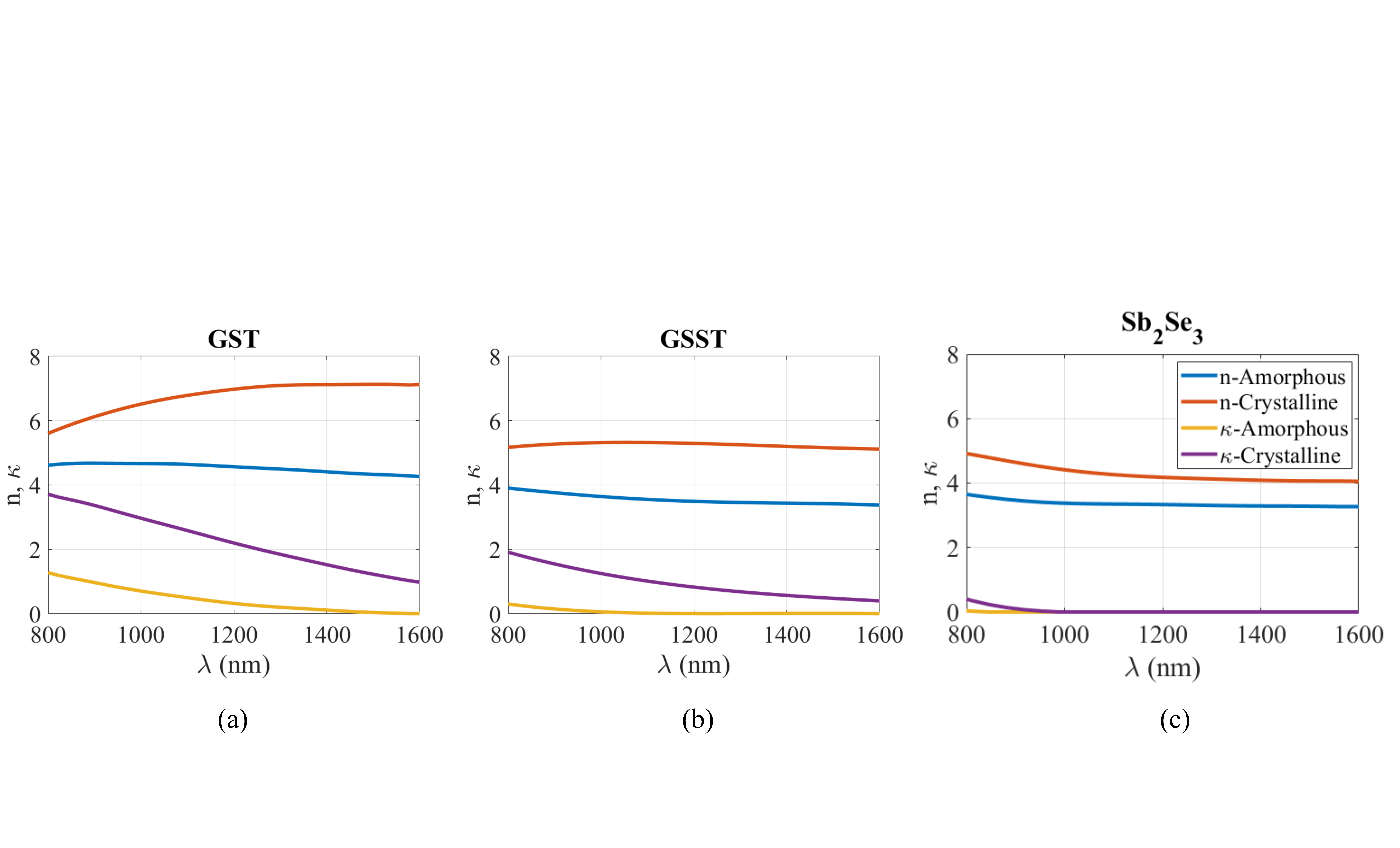}
     \vspace{-0.1in}
    \caption{Optical refractive index ($n$) and extinction coefficient ($\kappa$) for different PCM materials and wavelengths \cite{huang2023tunable, teo2022comparison}.}
    \label{Fig1_opt}
    \vspace{-0.1in}
\end{figure*}

With the increased complexity and growth in computationally expensive and data-driven applications, 
conventional CMOS-integrated circuits fail to meet the performance demands of emerging computing and communication systems \cite{shafiee2022silicon}. To address such demands, silicon photonics (SiPh) has emerged with a promise of ultra-fast communication and high-performance computation with enhanced energy efficiency \cite{mirza_charach, SiPh_codesign}. More recently, silicon photonic devices have been integrated with PCMs to enable reconfigurable photonic systems. The phase state of PCMs can change from amorphous to crystalline, and vice versa, upon a change in the material temperature when heated. This leads to a drastic nonvolatile change in the material's optical and electrical properties, making them promising candidates to realize photonic memory cells \cite{rios2015integrated, survey_shafiee}.

Conventional CMOS-based volatile memories, such as SRAMs and DRAMs, are reaching their limits in terms of energy efficiency, volume, and speed. This has motivated the need for new memory technologies, such as PCM-based photonic memories, to be used as the main memory or storage-class memory in future computing systems. Compared to other nonvolatile memories such as ReRAMs, PCM-based photonic memories can achieve higher scalability, energy-efficiency, stability, and bandwidth \cite{narayan2022architecting,kim2018future, feldmann2021parallel_nature_tensor_core}.

A PCM-based photonic memory cell can be obtained by depositing a PCM on top of a silicon-on-insulator (SOI) waveguide \cite{li2020experimental_SiN_Si}. Once the PCM is heated, its state can change from amorphous to crystalline, and vice versa. Overall, PCMs in the crystalline state show a higher refractive index than in the amorphous state \cite{rios2015integrated}. The refractive-index contrast between the crystalline and amorphous states will affect the optical transmission of the cell, which in turn can be used to store single or multiple bits per cell. Different optical transmission levels can be realized in PCM-based photonic memory cells by controlling the heat source, and hence the crystallization and amorphization dynamics of the cells (PCMs can be partially crystallized---i.e., intermediate state). This helps to store multiple bits per cell. Nevertheless, as the number of bits per cell increases, due to the need for a larger change in the optical transmission of the cell, a higher portion of the PCM must be crystallized, leading to higher set energies per cell \cite{survey_shafiee, rios2015integrated,rios2021ultra_phaseshifter,li2019fast_multilevel_5bit, thakkar2017dyphase}.


The novel contribution of this paper is in presenting a detailed design-space exploration for PCM-based photonic memories, from material to system level while using three well-known PCMs: GST, GSST, and Sb$_2$Se$_3$. Such an exploration helps realize the optimal cell design to meet specific energy-efficiency, footprint, and maximum-tolerable-loss requirements in PCM-based photonic memory cells. By modeling optical properties of the three PCMs, we perform multi-physics simulations (FDE, FDTD, and HEAT) for PCM-based photonic memory cells, to understand the impact of changing the PCM and waveguide geometries on cell-level performance (e.g., optical loss and power required for a PCM state change). In addition, we consider a memory array design from \cite{feldmann2019integrated} to show how the power consumption and scalability of a memory array changes when using different cells explored in this paper. 


\vspace{-0.1in}
\section{Background}\label{2_background} 
In this section, we present an overview of the fundamentals of PCMs and how PCM-based photonic memory cells work.

\vspace{-0.1in}
\subsection{Fundamentals and Properties of PCMs}
The state of a PCM can change from the amorphous to crystalline state, and vice versa, in a nonvolatile manner, leading to different optical and electrical properties.
One important material parameter of PCMs
is the melting temperature ($T_l$). The regions of the material after absorbing the energy from a heat source that have a temperature
above $T_l$ will be melted and quenched. The quenched region will have an amorphous structure
regardless of the initial state of the material. This process is
called \textit{reset}. Another important material parameter is the
crystallization temperature ($T_g$). When a PCM in the
intermediate state (or amorphous state) absorbs energy from a heat source,
the regions of the material with temperatures larger
than $T_g$ and lower than $T_l$ will recapture the crystalline structure. This process is called \textit{set}. Note that $T_l>T_g$, making the reset the most power-hungry procedure, compared to set.


 A PCM is in an intermediate state when a portion of the material has an amorphous state and the rest has a crystalline state. The crystallized area of the PCMs can be estimated by analyzing the temperature distribution of the cell when being heated \cite{youngblood2019tunable}. Temperature distribution in a PCM can be calculated by solving the unsteady-transient heat flow equation in the cell upon the heat transfer in the material \cite{li2019fast_multilevel_5bit}.
The required energy to trigger the phase transition of a PCM can be provided electrically, thermally, or optically \cite{narayan2022architecting}. For electrically (thermally) controlled PCMs, a PN junction (microheater) can be used to apply heat and initiate the phase transition \cite{rios2021ultra_phaseshifter}. When triggered optically, a laser pulse with specific power and duration will be used to set or reset the cells.  

 \begin{figure}[t]
    \centering
    \includegraphics[width=0.40\textwidth]{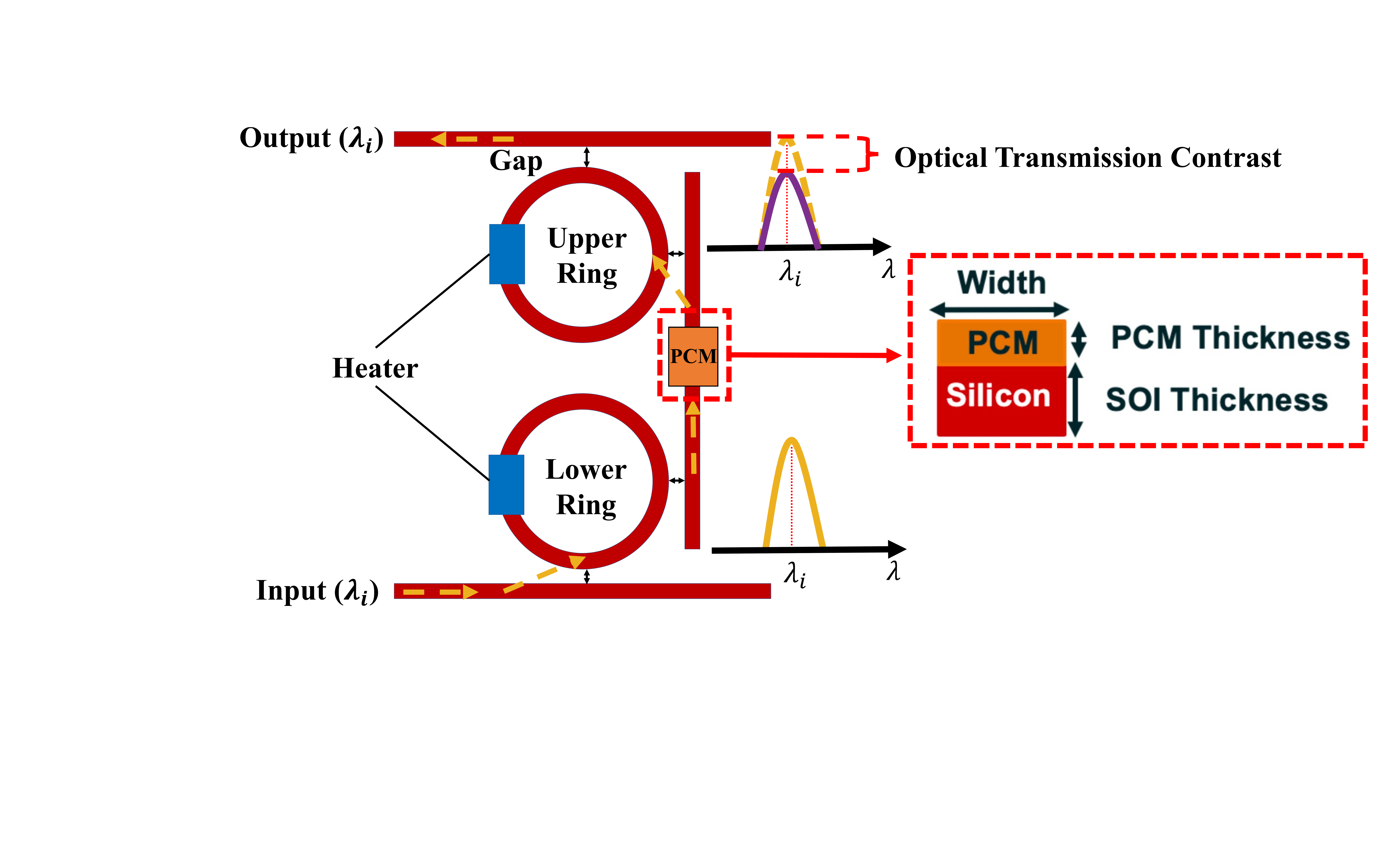}
    \caption{Unit cell of a PCM-based photonic memory \cite{feldmann2019integrated}.} \label{cell_arch}
    \vspace{-0.2in}
\end{figure}

To implement PCM-based photonic memory cells, understanding the optical properties of PCMs is important. Upon a phase (state) transition, the optical refractive index of the PCM, and hence the optical transmission of the cell, will change drastically. This can be used to store data on the cell's optical transmission levels. The optical refractive index profile of three PCMs (GST, GSST, and Sb$_2$Se$_3$) is shown in Fig. \ref{Fig1_opt}. Observe the drastic contrast between the crystalline and amorphous state of the PCMs. Note that for C-band (1530–1565 nm), GST shows the highest contrast in refractive index when shifting from amorphous to the crystalline state, and vice versa. This makes GST the most suitable candidate to implement PCM-based photonic memory cells. In addition, observe that the PCMs in the crystalline state have a much higher extinction coefficient compared to their amorphous state. This leads to higher absorption of the optical power in the crystalline state compared to the amorphous state. The absorbed optical power will be converted to heat, and can be used to trigger the phase transition in PCMs. 
To estimate the optical refractive index profile of the PCMs in any intermediate state, one can use the Lorenz model in \cite{wang2021scheme, survey_shafiee}.\vspace{-0.05in} 

\begin{figure}[t]
    \centering
    \includegraphics[width=0.3\textwidth]{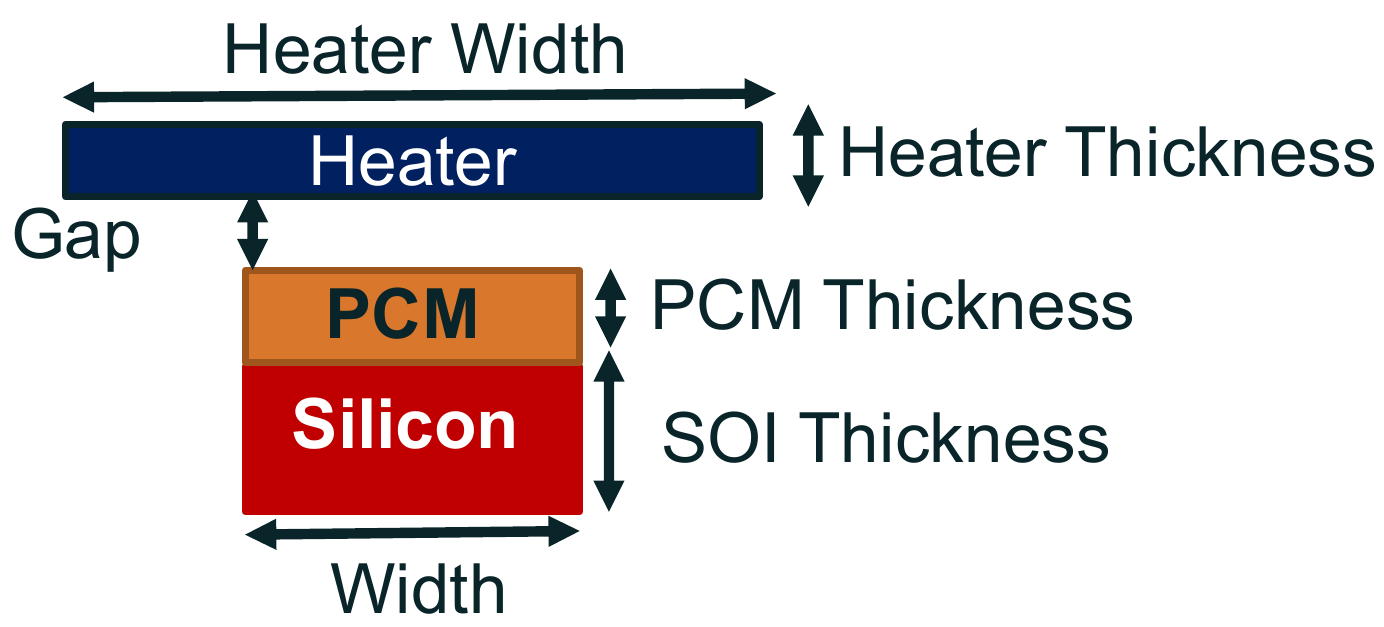}
    \caption{Designed PCM-based photonic memory cell. Set and reset are carried out using a microheater on top of the cell.}
    \label{cell_heater}
    \vspace{-0.23in}
\end{figure}


 
\begin{figure*}[t]
    \centering
    \includegraphics[width=0.9\textwidth]{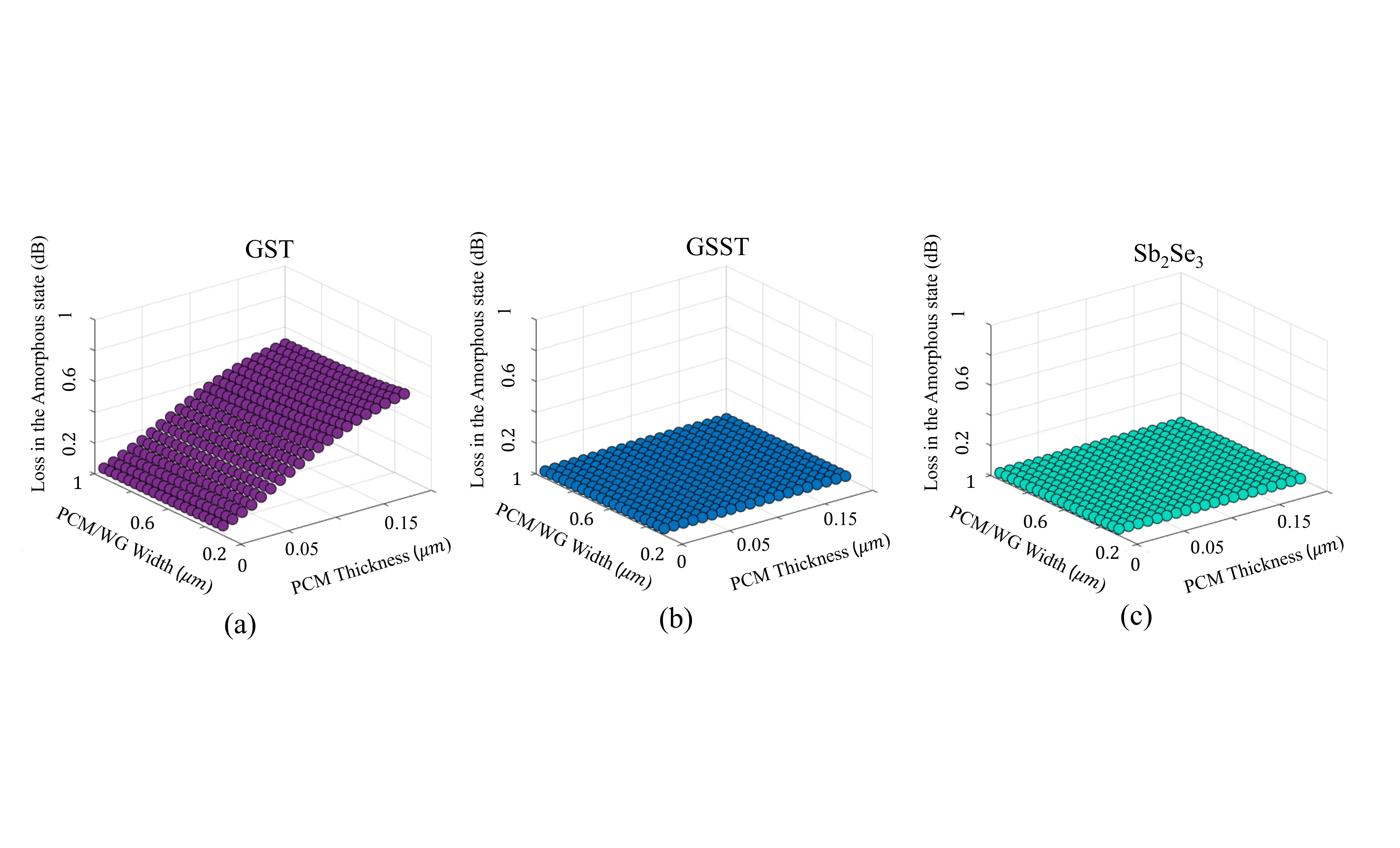}
     \vspace{-0.15in}
    \caption{Insertion loss of PCM-based photonic memory cells with different materials and geometries. WG: Waveguide.}
    \label{loss_amor}
   \vspace{-0.1in}
\end{figure*}

 \subsection{PCM-based Photonic Memory}
A PCM-based photonic memory cell can be realized by depositing a PCM on top of an SOI waveguide. The schematic of a PCM-based photonic memory cell is shown in Fig. \ref{cell_arch} \cite{feldmann2019integrated}. In this design, the light in the input waveguide couples to the lower ring, and then passes through the waveguide with the PCM on top of it (i.e., memory cell). The heaters on the rings are responsible for tuning the resonant wavelength in the rings
Note that in this design, both rings should have the same resonant wavelength to ensure correct operation with the same wavelength \cite{feldmann2019integrated}. 

 Because of the higher refractive index and extinction coefficient of PCMs in the crystalline state (see Fig. \ref{Fig1_opt}), as the PCM in the unit cell starts to crystallize, the optical transmission of the cell decreases due to the absorption of optical power in the PCM. The optical transmission contrast due to the absorption of light in the PCM helps realize multiple, distinct optical transmission levels between the initial and final state of the material, to store single or multiple bits per cell \cite{survey_shafiee, rios2015integrated}. As the optical transmission contrast between the initial and final state of the PCM increases, the cell is able to store a larger number of bits. However, this comes at a cost of higher power consumption or latency because a larger portion of the PCM needs to be crystallized. The light-PCM interaction reduces when using silicon (e.g., instead of silicon nitride \cite{rios2015integrated}) in PCM-based photonic memories \cite{li2020experimental_SiN_Si}. This makes the PCM-based photonic memories with SiPh more power-hungry with higher latency. However, PCM-based photonic memories with SiPh offer a more compact footprint, lower propagation loss, and compatibility with CMOS fabrication foundries \cite{li2020experimental_SiN_Si}.

 As mentioned earlier, phase transitions in PCMs can be triggered electrically, thermally, or optically. In this paper, the set and reset procedures are carried out thermally and using a microheater on top of the cell, due to the decreased light-matter interaction between silicon and PCM and, consequently, lower optical absorption in the PCM (on top of silicon waveguide) in amorphous and intermediate states. The schematic design of the cell with a microheater is shown in Fig. \ref{cell_heater}. Using this design, a low-power electrical signal with a long duration can be used to set (crystallize) the cell. A short-duration pulse with higher power (compared to the set pulse) can be used to reset (switching to the amorphous state) the cell, regardless of the initial state.
 \vspace{-0.1in}


\section{PCM-based photonic memory cells }\label{3_cells}
 In this section, we present a detailed design-space exploration of PCM-based photonic memory cells for the design demonstrated in Fig. \ref{cell_heater}, using GST, GSST, and Sb$_2$Se$_3$ PCMs. \vspace{-0.1in}

\subsection{Cell Insertion Loss}
 The insertion loss can be defined as the attenuation of the input optical signal when the cell is in an amorphous state. When a PCM is in the amorphous state (i.e., contains 0), it should not attenuate the input optical signal. The insertion loss originates from the extinction coefficient in the amorphous state (see Fig. \ref{Fig1_opt}). Note that as the cell starts to crystallize, the loss of the cell is in fact originating from the optical power absorption in the cell, which determines the optical transmission contrast being used to store data.

 Considering the cell in Fig. \ref{cell_heater}, Fig.\ref{loss_amor} shows the optical insertion loss for different PCMs of different geometries (width and thickness) at 1550~nm. Results are based on simulations in Lumerical MODE solver \cite{MODE}. Note that we consider the PCM's width and waveguide's width to be the same. Out of the three PCMs under test, GST shows the highest optical insertion loss in the amorphous state due to its higher extinction coefficient in the C-band (see Fig. \ref{Fig1_opt}(a)), where its loss can be as high as $\approx$0.6~dB/$\mu$m (see Fig. \ref{loss_amor}(a)). Moreover, note that the loss in the amorphous state for GST increases with its thickness, while the effect of PCM or waveguide width is insignificant. Despite GST's high insertion loss in the amorphous state, it has the highest contrast in the refractive index switching from the amorphous to crystalline state, making it the best candidate for photonic memories. Next are GSST and Sb$_2$Se$_3$ that are lossless in the amorphous state (see Figs. \ref{loss_amor}(b) and \ref{loss_amor}(c)),
 but compared to GST, have lower contrast in the refractive index between the two states. 
 Note that to realize PCM-based photonic memory cells, having low loss in the amorphous state and high refractive index contrast between crystalline and amorphous state is ideal.

 \begin{figure*}[t]
    \centering
    \includegraphics[width=0.9\textwidth]{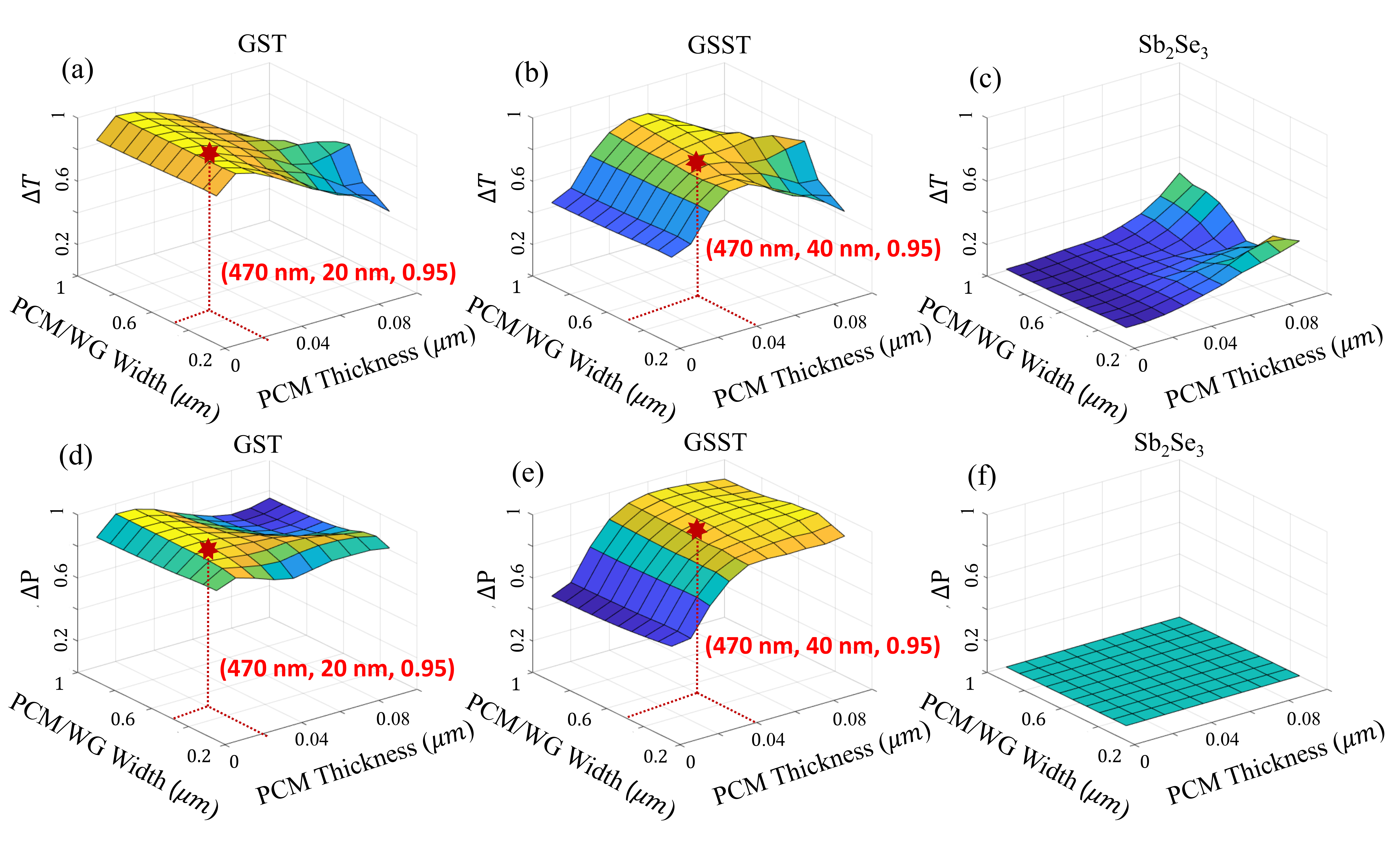}
     \vspace{-0.2in}
    \caption{(a)--(c) Optical transmission contrast ($\Delta T$) and (d)--(f) total absorption contrast ($\Delta P$) between crystalline and amorphous state for PCM-based photonic memory cells with GST, GSST, and Sb$_2$Se$_3$.  Simulations are based on Lumerical FDTD.}
    \label{transmission}
    \vspace{-0.15in}
\end{figure*}

\vspace{-0.1in}

\subsection{Cell Capacity}
 Cell capacity in PCM-based photonic memories is a parameter that can be determined by capturing the optical transmission contrast between the amorphous and partially or fully crystallized state of the PCM \cite{li2019fast_multilevel_5bit}. As the crystallization fraction increases, the optical-transmission changes increase due to increased attenuation of the input optical signal. This leads to a higher number of separable signal levels to store data, and hence storing a larger number of bits. For example, for a 2-bit PCM-based photonic memory cell, only 4 signal levels are needed to store data (00, 10, 01, 11). 
 
 The optical transmission contrast ($\Delta T$) and optical absorption contrast ($\Delta P$) between fully crystalline and fully amorphous state for 2-$\mu$m-long PCM-based photonic memory cells of different geometries and materials are shown in Fig. \ref{transmission}. Note that $\Delta T$ is not only a function of $\Delta P$ in the cells. $\Delta T$ partially originates from the optical-refractive-index mismatch between the PCM and SOI waveguide. The effect of the refractive-index contrast is more observable in Sb$_2$Se$_3$. We can see from Figs. \ref{transmission}(c) and \ref{transmission}(f) that for Sb$_2$Se$_3$, although $\Delta P$ is zero, the material unexpectedly shows some $\Delta T$ between the two states, which stems from the optical-refractive-index mismatch.  Note that such a $\Delta T$ in Sb$_2$Se$_3$ cannot be controlled actively as it is independent of the material absorption (or phase change of the material). In addition, Sb$_2$Se$_3$ shows lower refractive index contrast, and hence significantly lower $\Delta T$ compared to GST and GSST (see Fig. \ref{Fig1_opt}). These make Sb$_2$Se$_3$ not an ideal candidate to implement PCM-based photonic memories with SOI waveguides, necessitating some additional design optimization to address the optical-refractive-index mismatch.
 

To avoid optical-refractive-index mismatch when designing GST- and GSST-based photonic memory cells, one should pick a design where both $\Delta T$ and $\Delta P$ are maximum. Doing so ensures that the $\Delta T$ is stemming from the optical power absorption. Accordingly, considering Figs. \ref{transmission}(a) and \ref{transmission}(d), for a 2-$\mu$m-long GST cell, $\Delta T$ and $\Delta P$ are at 95\% when the thickness of the cell is about 20~nm with the width of 470~nm. Note that the impact of waveguide/PCM width on $\Delta T$ and $\Delta P$ is negligible. This cell can store up to 6 bits (up to 64 separable signal levels), considering a $\approx$1\% ($0.96/64$) margin between each state of the cell. However, this cell suffers from 0.2~dB/$\mu$m insertion loss in the amorphous state (see Fig. \ref{loss_amor}(a)). Using the same approach, we can design a 2-$\mu$m-long, 40-nm thick GSST-based cell with a width of 470~nm to store 6 bits per cell, but with no insertion loss in the amorphous state.

 The bit capacity of a cell is determined by adjusting the crystallized fraction of the cell. As mentioned in Section \ref{2_background}, the refractive index of a PCM in an intermediate state can be estimated using the Lorenz model from \cite{wang2021scheme, survey_shafiee}, and assuming a uniform phase transition in the PCM's volume from amorphous to crystalline state. Using the Lorenz model and FDTD simulations, and assuming $\approx$1\% margin to separate transmission levels \cite{li2019fast_multilevel_5bit}, to store a maximum of 2 (4) bits per cell, we found that up to 20\% (40\%) of the PCM needs to be crystallized when using GST and GSST. To store 6 bits per cell, these cells should be fully crystallized. Note that the aforementioned values are the required crystallization fraction for the extreme cases for writing "2$^n$$-$1" ($n$ is the bit capacity of the cell) to the cells. The crystalline fraction can be controlled by tuning the power and duration of the heat source being used to set the cells, and, as it was mentioned, it can be estimated by solving the unsteady-transient heat transfer equation in the PCM's volume \cite{wang2021scheme}. The reset procedure will be the same regardless of the maximum number of bits stored, as the PCM should return to its initial amorphous state. Storing multiple bits per a PCM-based photonic memory cell can be challenging due to the essential need for more complex programming and detection policies at the architectural level, when scaling the cells to implement memory arrays.\vspace{-0.1in}
 
\subsection{Latency and Power Consumption}
The latency and power consumption of a PCM-based photonic memory cell are a function of the cell's maximum bit capacity. As the cell's bit capacity increases, due to the need for a higher transmission contrast, more energy is required to reach higher levels of crystallization. Using the design in Fig. \ref{cell_heater}, a microheater is designed to set and reset PCM-based photonic memory cells. The heater material is Ti/TiN with $\rho=$~60~$\mu.\Omega$.cm and a sheet resistance of 5.5~$\Omega$/sq. The melting temperature of the heater material (Ti/TiN) is 1941 K, considered to avoid melting the heater upon heating the PCM. The thickness of the heater is 110~nm with the width and length of 2~$\mu$m, placed 600~nm above the waveguide to reduce metallic absorption due to metal-light interaction. Lumerical HEAT \cite{MODE} is used to carry out unsteady-transient heat transfer simulations to capture the temperature distribution in the PCMs (only GST and GSST; see Section 3.2), as a function of exposure time for a given electric power applied to the heater. 

\begin{figure}[t]
    \centering
    \includegraphics[width=0.48\textwidth]{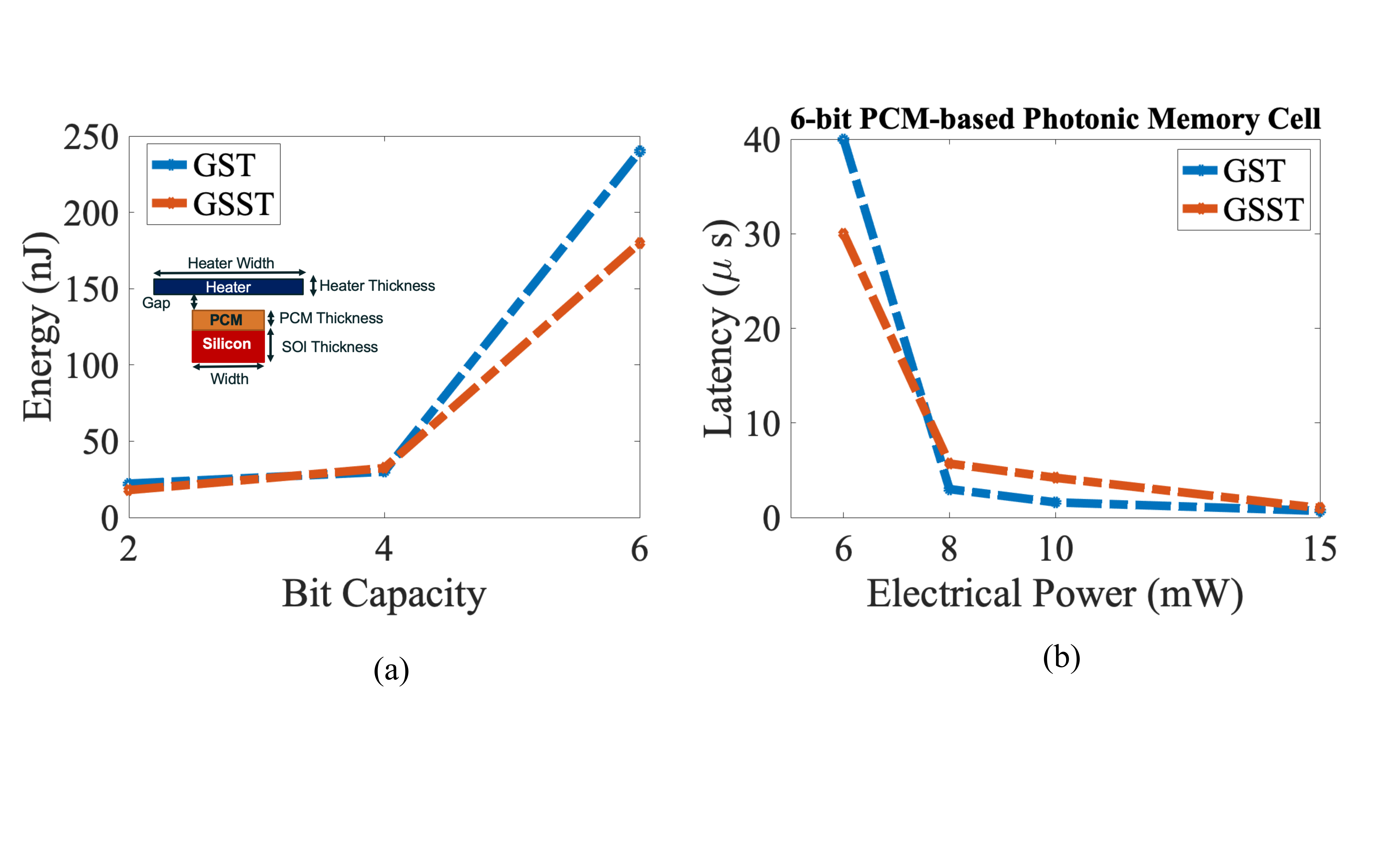}
     \vspace{-0.25in}
    \caption{(a) Maximum set energy versus cell's bit capacity for two PCM-based photonic memory cells using GST and GSST. A constant 6~mW electrical pulse with various pulse widths is used. (b) Power-latency trade-off for the same cells to achieve fully crystalline state to write "111111" to the cell.}
    \label{latency}
    \vspace{-0.22in}
\end{figure}

Fig.\ref{latency}(a) shows the maximum set energy for the GST- and GSST-based photonic memory cells designed in Section 3.2,  when a 6~mW electrical pulse is applied to the heater with different pulses ($E=P.t$, where $t$ is the pulse duration and $P$ is the electric power applied to the heater). The $T_g$ for GST and GSST is considered to be 453~K and 423~K, respectively. In addition, the melting temperature of 890~K and 900~K is considered for GST and GSST, respectively \cite{aryana2021suppressed, li2019fast_multilevel_5bit, survey_shafiee, GSST_melting}. We can see from Fig.\ref{latency}(a) that, in general, as we increase the cell's bit capacity, the maximum energy required to set the cell (energy that is required to write "2$^n$$-$1", where $n$ is the cell's bit capacity) also increases. This is due to the essential need for larger transmission and optical absorption contrast. For example, in a 6-bit PCM-based photonic memory cell using GSST, the maximum energy of 175~nJ is required to write "111111" to the cell, while for GST, this energy can be as high as 248~nJ. An electric pulse of 40~mW with a duration of 3.5~$\mu$s is used to reset the cells by reaching the melting temperature of the PCMs, hence returning to the amorphous state. 

The power-latency trade-off for the 6-bit GST- and GSST-based cells is shown in Fig. \ref{latency}(b). As it can be seen, as we increase the maximum set power, the latency decreases and the trend is nonlinear. In addition, note that for power values lower than 6~mW, phase transition cannot be triggered, regardless of pulse duration. In other words, the required energy to trigger a specific phase transition is not always the same, and it depends on the electrical power used to write on the cells. This effect stems from the physical mechanism of the heat transfer from the heater to the PCMs given the sample's thermal properties, such as thermal conductivity, specific heat capacity, and density.

\section{PCM-based Memory Arrays}\label{4_arch}
\begin{figure}[t]
    \centering
    \includegraphics[width=0.49\textwidth]{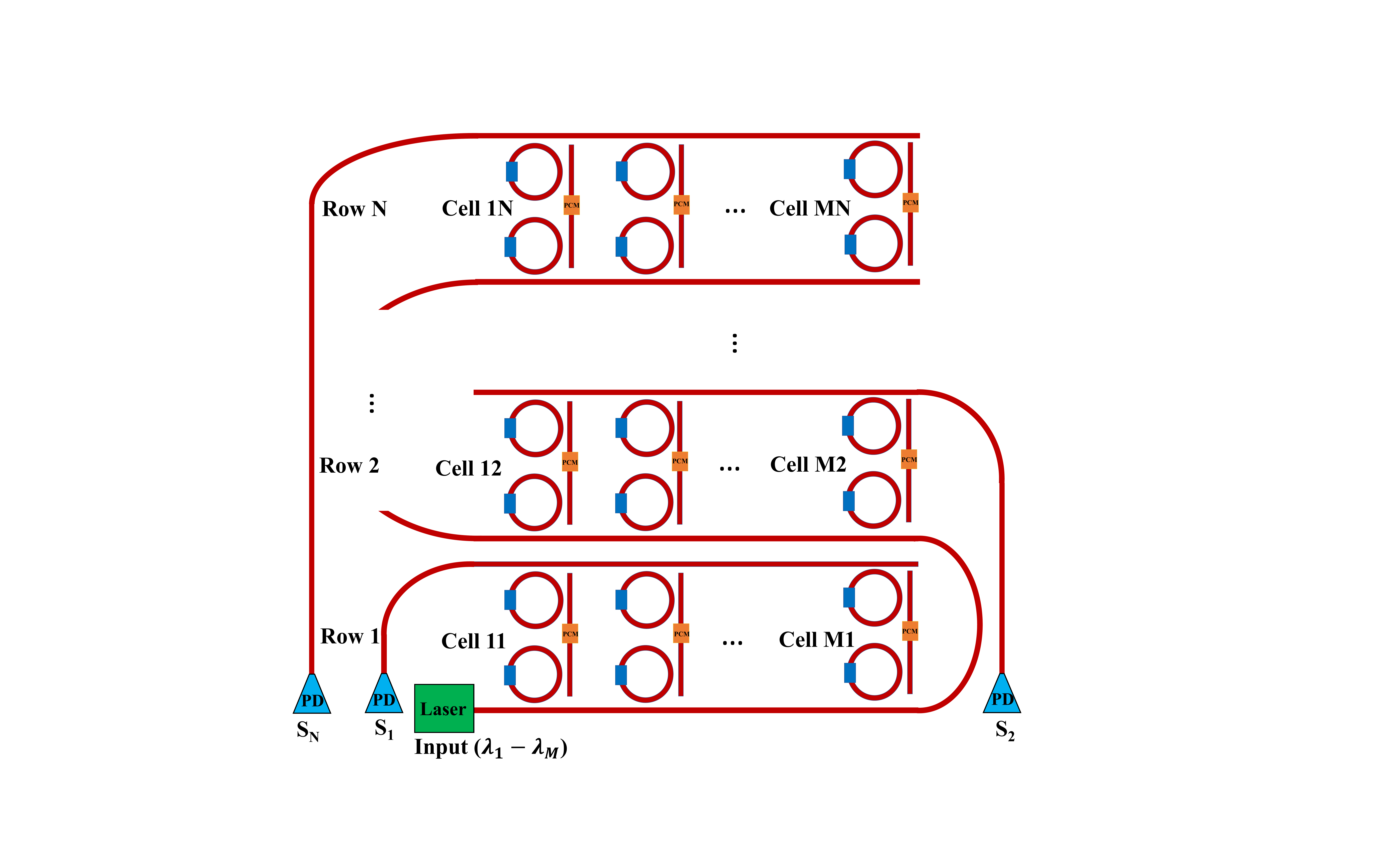}
     \vspace{-0.15in}
    \caption{PCM-based photonic memory array based on the design presented in \cite{feldmann2019integrated}. PD: Photodetector.}
    \label{arch_main}
    \vspace{-0.1in}
\end{figure}

Leveraging the cell introduced in Section \ref{2_background} (see Fig. \ref{cell_arch}), one can realize a PCM-based photonic memory array by cascading $M$ cells per row, and for the total number of $N$ rows with the configuration depicted in Fig. \ref{arch_main}. Here, we consider the design presented in \cite{feldmann2019integrated}. Note that the original design in \cite{feldmann2019integrated} used different ring radii to induce different resonant wavelength shifts in a row, while in our work microheaters on the rings are used to realize the required resonant shift to read and write data with the PCM-based photonic memory cells. The reason for using heaters instead of different radii is to actively control the resonant shift in the rings and the spacing between the resonant peaks, which creates an additional degree of freedom when designing a memory array. Due to using heaters for tuning the rings in this design, there is no need for fine-tuning the input, drop, and output gaps in the rings associated with each PCM cell. The resonant wavelength of the rings in each row is slightly different ($\Delta \lambda$ = 850 pm \cite{feldmann2019integrated}), which is controlled by the heaters in our design.
The readout of each cell in the memory depicted in Fig.~\ref{arch_main} can be done in two steps. First, we can select the row to be read using output ports S$_1$ to S$_N$. Then, the cell to be read from within each row can be selected via the input wavelength,
due to the slight difference between the resonant wavelengths of all rings in a row \cite{feldmann2019integrated}. Finally, the optical signal transmission from each cell can be converted to an electrical signal via photodetectors (PDs) at the end of each output port, to retrieve the stored data. 

\begin{figure}[t]
    \centering
    \includegraphics[width=0.47\textwidth]{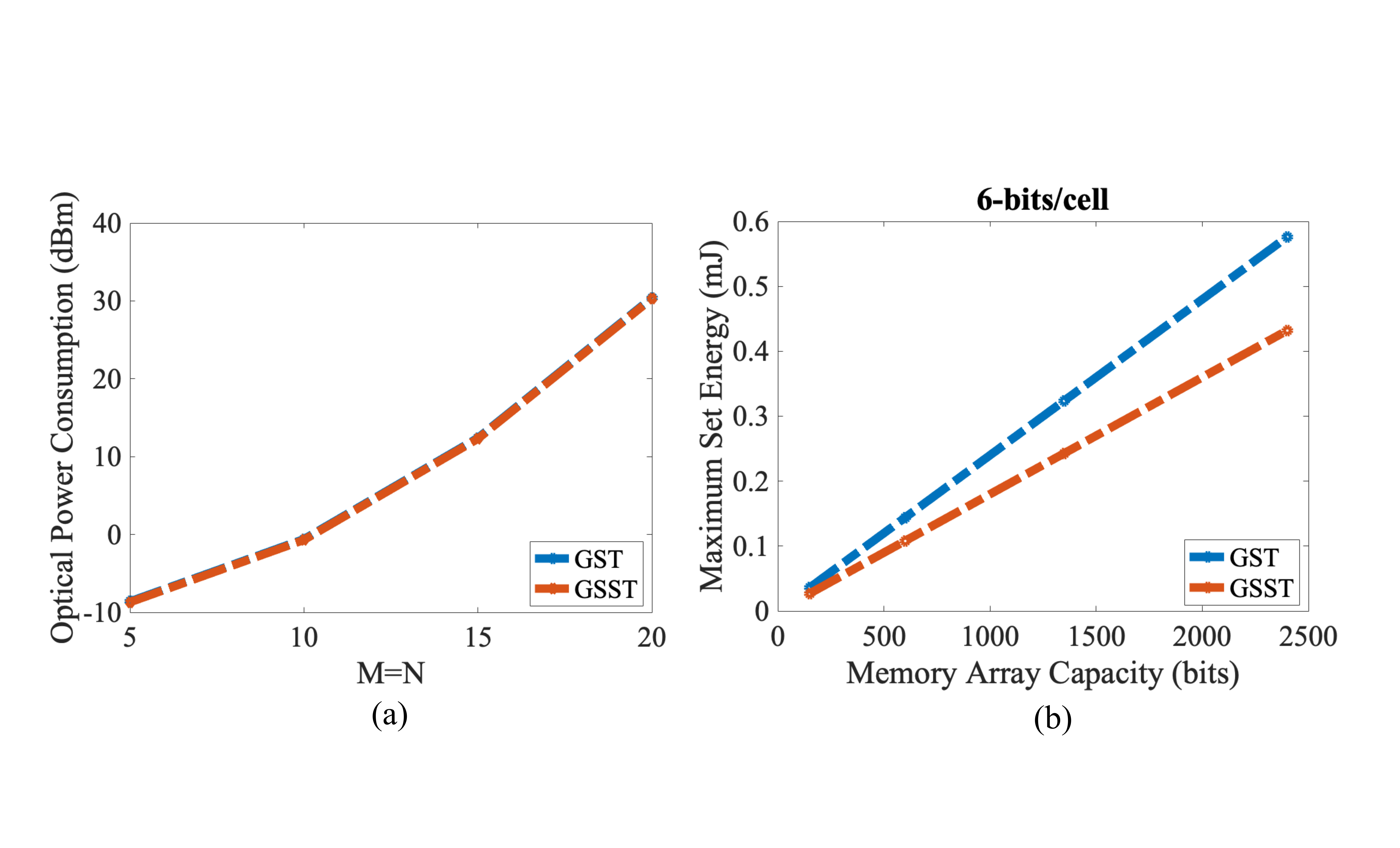}
     \vspace{-0.1in}
    \caption{(a) Optical power consumption and (b) maximum set energy (6-bits per cell) for the memory array in Fig. \ref{arch_main}.}
    \label{laser_penalty}
    \vspace{-0.1in}
\end{figure}

Employing the two cells designed in Section \ref{3_cells}, we explore the scalability of the memory array in Fig. \ref{arch_main}. The required laser output optical power ($P_{lsr}$) for reading from the last cell in each row in this memory array can be defined as \cite{shafiee_loci}:
\begin{equation}
     P_{lsr} \geq S_{PD} + [(N\times M)-1 + (M-1)] \times L_{p} + 2L_{d} + L_{amorphous}(\lambda).\label{penalty}
\end{equation}
Assuming $M=N$, and that all the cells contain 0 (amorphous state), PD's sensitivity of $S_{PD}=$~$-$11.7~dBm \cite{9199100PD, palmieri2020enhanced}, average ring passing/drop loss of $L_{p} =L_d=$~0.1~dB \cite{feldmann2019integrated}, and average loss in the amorphous state (for the wavelength range of $\lambda=$~1530--1565~nm) of $L_{amorphous} =$~0.35~dB for GST, and 0~dB for GSST, the required laser power to read from the last cell in the last row of the memory array (i.e., worst-case optical power consumption) is shown in Fig. \ref{laser_penalty}(a). The insertion loss in the amorphous state is considered constant due to the small spectrum spacing between the wavelengths used to read the data. Note that too much increase in the readout optical power requirement can lead to an increased number of read errors due to the change in the state of the PCM (especially for the case when the cells in the first row of the memory array have a high crystallization fraction). We can see from the results in Fig. \ref{laser_penalty}(a) that the array size plays an important role and as we increase the memory array size, the optical power at the input increases dramatically and can be as high as 30.4~dBm for a PCM-based photonic memory, regardless of the maximum number of bits per cell. Note that the effect of loss in the amorphous state is insignificant in the memory design presented in Fig. \ref{arch_main}. This is because of using different wavelengths to read different cells in a row. Therefore, each read optical signal experiences loss from a single PCM cell it is reading from, and losses from other cells will not impact this signal. However, the read signal suffers from higher optical losses as the memory array scales up, due to the increase in the number of rings it passes (and higher propagation loss, not discussed in this paper).

The maximum set energy is another parameter that changes with scaling of the memory array capacity. The maximum set energy for the memory array using 6-bit GST- and GSST-based cells to write "111111" is shown in Fig. \ref{laser_penalty}(b) for different memory array capacities (i.e., total number of bits stored) with $M=N$. As can be seen, as we increase the array capacity, the maximum write energy of the entire memory increases linearly, and it can be as high as 0.6~mJ for GST and 0.4~mJ for GSST. Note that a 6~mW electrical pulse is used to write on the cells via heaters. Considering the results in Fig. \ref{laser_penalty}, we can see that scaling up a memory array to increase its capacity is infeasible without further optimization of the cell's structure due to high input optical power required to compensate for the losses. For example, to store 2400~bits (120-bits per row and 6-bits per cell when $M=N=$20), the input laser should provide at least 30.4~dBm to compensate for the losses throughout the memory array. Consequently, the 1\% margin between optical transmission levels considered in this paper may lead to unreliable readouts due to the undesired change in the state of the cells due to increased input optical power.
This motivates the need for a trade-off between the transmission level's margin and the number of levels, and therefore the cell's bit capacity. Moreover, optical transmission drift is another limitation that can lead to unreliable readout of cells. Such drifts impose a higher set pulse duration and lower bit capacity (to achieve a larger margin between the optical transmission levels) to stabilize the cell's state when writing the data \cite{survey_shafiee, li2019fast_multilevel_5bit}. Another limiting factor in scaling the design in Fig. \ref{arch_main} is the free-spectral range (FSR) of the rings. We cannot arbitrarily increase the number of rings to store more bits per array. Increasing the number of rings per row necessitates a larger number of operating wavelengths to store and read the data from the cells. This leads to the essential need for rings with a large FSR, which requires smaller rings (FSR is inversely proportional to ring radius) at the cost of increased optical loss in the rings.

\section{Conclusion} \label{5-conclusion}
In this paper, we presented a design-space exploration of PCM-based photonic memories with silicon photonics using three well-known PCMs, namely GST, GSST, and Sb$_2$Se$_3$. Parameters such as optical insertion loss of the cell in the amorphous state, optical transmission contrast between amorphous and crystalline state, cell's bit capacity, and set and reset energies are explored to design an optimized photonic memory cell. We showed that for thermally controlled PCM-based photonic memory cells with GST or GSST, as the bit capacity of the cells increases, the maximum set energy also increases drastically. Finally, we presented an example of a memory array using the optimized memory cells and explored the scalability and maximum set energy in the array as the size of the array changes. Our results show the promise of PCM-based photonic memories and the critical need for cross-layer design co-optimization (material to array level) to minimize energy and latency costs in such memories.

\section*{ACKNOWLEDGEMENTS}
This work was supported in part by the National Science Foundation under grants CCF-2006788 and CNS-2046226.
\vspace{-0.05in}


\bibliographystyle{IEEEtran}
\bibliography{IEEEabrv,main}

\begin{thebibliography}{10}
\providecommand{\url}[1]{#1}
\csname url@samestyle\endcsname
\providecommand{\newblock}{\relax}
\providecommand{\bibinfo}[2]{#2}
\providecommand{\BIBentrySTDinterwordspacing}{\spaceskip=0pt\relax}
\providecommand{\BIBentryALTinterwordstretchfactor}{4}
\providecommand{\BIBentryALTinterwordspacing}{\spaceskip=\fontdimen2\font plus
\BIBentryALTinterwordstretchfactor\fontdimen3\font minus
  \fontdimen4\font\relax}
\providecommand{\BIBforeignlanguage}[2]{{%
\expandafter\ifx\csname l@#1\endcsname\relax
\typeout{** WARNING: IEEEtran.bst: No hyphenation pattern has been}%
\typeout{** loaded for the language `#1'. Using the pattern for}%
\typeout{** the default language instead.}%
\else
\language=\csname l@#1\endcsname
\fi
#2}}
\providecommand{\BIBdecl}{\relax}
\BIBdecl

\bibitem{huang2023tunable}
Y.-S. Huang \emph{et~al.}, ``Tunable structural transmissive color in
  fano-resonant optical coatings employing phase-change materials,''
  \emph{arXiv preprint arXiv:2302.03207}, 2023.

\bibitem{teo2022comparison}
T.~Y. Teo \emph{et~al.}, ``Comparison and analysis of phase change
  materials-based reconfigurable silicon photonic directional couplers,''
  \emph{Opt. Mater. Express}, vol.~12, no.~2, pp. 606--621, 2022.

\bibitem{shafiee2022silicon}
A.~Shafiee \emph{et~al.}, ``Silicon photonics for future computing systems,''
  \emph{Wiley Encyclopedia of Electrical and Electronics Engineering}, pp.
  1--26, 2022.

\bibitem{mirza_charach}
A.~Mirza \emph{et~al.}, ``Characterization and optimization of coherent
  mzi-based nanophotonic neural networks under fabrication non-uniformity,''
  \emph{IEEE Transactions on Nanotechnology}, vol.~21, pp. 763--771, 2022.

\bibitem{SiPh_codesign}
Q.~Cheng \emph{et~al.}, ``Silicon photonics codesign for deep learning,''
  \emph{Proc. IEEE}, vol. 108, no.~8, pp. 1261--1282, 2020.

\bibitem{rios2015integrated}
C.~R{\'\i}os \emph{et~al.}, ``Integrated all-photonic non-volatile multi-level
  memory,'' \emph{Nature photonics}, vol.~9, no.~11, pp. 725--732, 2015.

\bibitem{survey_shafiee}
A.~Shafiee \emph{et~al.}, ``A survey on optical phase-change memory: The
  promise and challenges,'' \emph{IEEE Access}, vol.~11, pp. 11\,781--11\,803,
  2023.

\bibitem{narayan2022architecting}
A.~Narayan \emph{et~al.}, ``Architecting optically controlled phase change
  memory,'' \emph{ACM Transactions on Architecture and Code Optimization},
  vol.~19, no.~4, pp. 1--26, 2022.

\bibitem{kim2018future}
S.~K. Kim and M.~Popovici, ``Future of dynamic random-access memory as main
  memory,'' \emph{MRS Bulletin}, vol.~43, no.~5, pp. 334--339, 2018.

\bibitem{feldmann2021parallel_nature_tensor_core}
J.~Feldmann \emph{et~al.}, ``Parallel convolutional processing using an
  integrated photonic tensor core,'' \emph{Nature}, vol. 589, no. 7840, pp.
  52--58, 2021.

\bibitem{li2020experimental_SiN_Si}
X.~Li \emph{et~al.}, ``Experimental investigation of silicon and silicon
  nitride platforms for phase-change photonic in-memory computing,''
  \emph{Optica}, vol.~7, no.~3, pp. 218--225, 2020.

\bibitem{rios2021ultra_phaseshifter}
C.~Rios \emph{et~al.}, ``Ultra-compact nonvolatile photonics based on
  electrically reprogrammable transparent phase change materials,'' \emph{arXiv
  preprint arXiv:2105.06010}, 2021.

\bibitem{li2019fast_multilevel_5bit}
X.~Li \emph{et~al.}, ``Fast and reliable storage using a 5 bit, nonvolatile
  photonic memory cell,'' \emph{Optica}, vol.~6, no.~1, pp. 1--6, 2019.

\bibitem{thakkar2017dyphase}
I.~G. Thakkar \emph{et~al.}, ``Dyphase: A dynamic phase change memory
  architecture with symmetric write latency and restorable endurance,''
  \emph{IEEE Transactions on Computer-Aided Design of Integrated Circuits and
  Systems}, vol.~37, no.~9, pp. 1760--1773, 2017.

\bibitem{feldmann2019integrated}
J.~Feldmann \emph{et~al.}, ``Integrated 256 cell photonic phase-change memory
  with 512-bit capacity,'' \emph{IEEE Journal of Selected Topics in Quantum
  Electronics}, vol.~26, no.~2, pp. 1--7, 2019.

\bibitem{youngblood2019tunable}
N.~Youngblood \emph{et~al.}, ``Tunable volatility of {Ge2Sb2Te5} in integrated
  photonics,'' \emph{Advanced Functional Materials}, vol.~29, no.~11, p.
  1807571, 2019.

\bibitem{wang2021scheme}
Y.~Wang \emph{et~al.}, ``A scheme for simulating multi-level phase change
  photonics materials,'' \emph{npj Computational Materials}, vol.~7, no.~1, p.
  183, 2021.

\bibitem{MODE}
\BIBentryALTinterwordspacing
{Ansys Lumerical}. [Online]. Available:
  \url{https://www.lumerical.com/products/}
\BIBentrySTDinterwordspacing

\bibitem{aryana2021suppressed}
K.~Aryana \emph{et~al.}, ``Suppressed electronic contribution in thermal
  conductivity of {Ge2Sb2Se4Te},'' \emph{Nature Communications}, vol.~12,
  no.~1, p. 7187, 2021.

\bibitem{GSST_melting}
F.~De~Leonardis \emph{et~al.}, ``Broadband electro-optical crossbar switches
  using low-loss {Ge2Sb2Se4Te1} phase change material,'' \emph{Journal of
  Lightwave Technology}, vol.~37, no.~13, pp. 3183--3191, 2019.

\bibitem{shafiee_loci}
\BIBentryALTinterwordspacing
A.~Shafiee \emph{et~al.}, ``Loci: An analysis of the impact of optical loss and
  crosstalk noise in integrated silicon-photonic neural networks,'' in
  \emph{Proceedings of the Great Lakes Symposium on VLSI 2022}, ser. GLSVLSI
  '22.\hskip 1em plus 0.5em minus 0.4em\relax New York, NY, USA: Association
  for Computing Machinery, 2022, p. 351–355. [Online]. Available:
  \url{https://doi.org/10.1145/3526241.3530365}
\BIBentrySTDinterwordspacing

\bibitem{9199100PD}
X.~Bi \emph{et~al.}, ``High sensitivity and dynamic-range 25 {Gbaud} silicon
  receiver chipset with current-controlled {DC} adjustment path and cube-shape
  {Ge-on-Si} {PD},'' \emph{IEEE TCSI: Regular Papers}, vol.~67, no.~11, pp.
  3991--4001, 2020.

\bibitem{palmieri2020enhanced}
A.~Palmieri \emph{et~al.}, ``Enhanced dynamic properties of {Ge-on-Si}
  mode-evolution waveguide photodetectors,'' in \emph{2020 International
  Conference on Numerical Simulation of Optoelectronic Devices (NUSOD)}.\hskip
  1em plus 0.5em minus 0.4em\relax IEEE, 2020, pp. 1--2.

\end{thebibliography}
%
%
\end{document}